\documentstyle[12pt]{article}
\def\p{\partial}
\def\d{\delta}
\def\tr{\,{\rm tr}\,}
\def\pr{\psi_{r}}
\def\prt{\psi_{r}^{T}}
\def\phrt{\phi_{r}^{T}}
\def\prbt{\bar{\psi}_{r}^{T}}

\def\prb{\bar{\psi}_{r}}
\def\aor{\alpha_{1}^{\rm ort}}
\def\am{A_{\mu}}
\def\amt{A_{\mu}^{T}}
\def\an{A_{\nu}}
\def\ant{A_{\nu}^{T}}
\def\al{A_{\lambda}}
\def\as{A_{\sigma}}
\def\alt{A_{\lambda}^{T}}
\def\ast{A_{\sigma}^{T}}
\def\ls{i\bar{\psi}\gamma^\mu(\partial_\mu +\am)\psi}
\def\phr{\phi_{r}}
\def\la{\lambda^{a}}
\def\lb{\lambda^{b}}

\def\phrb{\bar{\phi}_{r}}
\def\be{\begin{equation}}
\def\ee{\end{equation}}
\def\bea{\begin{eqnarray}}
\def\eea{\end{eqnarray}}
\def\wz{Wess-Zumino }
\def\e{\epsilon}
\def\a{\alpha}
\vspace{5cm}

\title{  \hfill{SMI-35-94}
\\SO(N) invariant \wz action and its quantization}
\vspace{.7cm}
\author{ \mbox{}
\\ S.A.Frolov\thanks{Alexander von Humboldt fellow}
\mbox{} \\ \vspace{0.4cm} Section Physik, Munich University
\vspace{-0.5cm} \mbox{} \\ Theresienstr.37, 80333 Munich, Germany
\thanks{Permanent address:\ Steklov Mathematical Institute, Moscow}
\mbox{} \\ \\ \vspace{.6cm} A.A.Slavnov and C.Sochichiu\thanks{and Moscow
State University}
\vspace{-0.5cm} \mbox{} \\ Steklov Mathematical Institute
\vspace{-0.1cm} \mbox{}  \\ Vavilov st.42, GSP-1, 117966 Moscow, RUSSIA
\date{}}

\begin{document}
\maketitle
\vspace{4.5cm}
\begin{abstract}
A consistent quantization procedure of anomalous chiral models is
discussed. It is based on the modification of  the classical action by
adding Wess-Zumino terms. The $SO(3)$ invariant WZ action for the $SO(3)$
model is constructed. Quantization of the corresponding modified theory is
considered in details.
\end{abstract}

 \section{Introduction}

In this paper we consider a possibility of consistent canonical
quantization of anomalous gauge theories based on the modification of the
classical action by adding the Wess-Zumino (WZ) term. It is known that the
straightforward quantization of anomalous models leads to unconsistent
theory breaking either unitarity or Lorentz covariance or
renormalizability \cite{a, b, c}. Usually the absence of anomalies is
considered as a criterion of choosing physically acceptable model.
However there exists another point of view according to which the
appearance of anomalies leads to increasing of the number of degrees of
freedom: some excitations which can be eliminated  in the classical theory
by gauge transformation become physical. This point of view was adopted by
Polyakov \cite{1} in the relativistic  string model, by Jackiw and
Rajaraman \cite{2} in the chiral Schwinger model and by Faddeev and
Shatashvili \cite{3} in nonabelian  anomalous chiral models. However at
present it is not yet clear whether this point of view may lead to a
consistent theory. A problem arises  already in the process of quantizing
anomalous models. As was pointed out by Faddeev \cite{4} in anomalous
models some classical first class constraints transform into the second
class constraints. For example, if one quantizes the theory in the
temporal gauge $ A_0=0 $ one should select the physical subspace by
imposing the condition $ \varphi |\psi> =0 $ where $ \phi $ is a Gauss law.
In nonanomalous models $ \varphi $ form a set of first class constraints
generating the gauge transformations.  However in the presence of anomaly
the algebra of consraints is modified. In the commutation relations the
Schwinger term arises which has a meaninng of a 2-cocycle on the gauge
group \cite{4, 5, 6}. That means the conditions $ \phi |\psi> =0 $ are
inconsistent and the quantization procedure should be revised.

The origin of appearence of anomalies is impossibility to introduce for
anomalous models a gauge invariant intermediate regularization which is
necessary to give a precise mathematical meaning to the quantum theory.
Any regularization changes the type of constraints transforming some of
the first class constraints to the second class ones. It suggests that one
possibility to perform a consistent quatization is to introduce a
Lagrangian regularization on the classical level and apply the canonical
quantization to the regularized theory. An alternative way is to introduce
a regularization in the framework of BRST quantization scheme \cite{5a}.
The first procedure was succefully applied to the abelian chiral models
and to the string model in refs.\cite{6', 7, 8}. As there is no gauge
invariant regularization, the resulting theory is not gauge invariant and
depends crucially on the particular regularization chosen. For example in
the two dimensional chiral Schwinger model some regularizations lead to
the consistent theory with  the positive definite Hamiltonian while the
others generate non physical ghost states. Although the resulting theory
in this case is not gauge invariant the invariance may be easely restored
by introducing a new field $g$ with values in the gauge group. In the case
of chiral gauge theories which we shall consider below it can be done by
making a gauge transformation $A_\mu \rightarrow A_\mu^g$; $\psi
\rightarrow \psi^g$ and considering $g$ as a new variable with the
transformation law $g \rightarrow h^{-1}g $. Taking into account that the
classical action is gauge invariant one sees that the dependence on the
chiral field $g$ enters only via regularizing fields. Integrating over all
the fields except for $A_\mu $ and $g$ one gets an effective action for
these fields. The part of this action which depends on $g$ appears due to
the gauge noninvariance of the regularization and is a 1-cocycle on the
gauge group. When the regularization parameter goes to infinity this
part reduces to the WZ action \cite{9}. This action may be calculated also
by the algebraic and geometric methods \cite{4, 5, 10, 11}. Its gauge
variation gives the anomaly  which is therefore an infinitesimal
1-cocycle. It suggests another possibility of quantizing anomalous
theories.  One can modify the classical gauge invariant action by adding
the WZ term and quantize the modified action \cite{3}. The modified
classical action is not gauge invariant however one can quantize it by
imposing some gauge condition and then prove that the quantum
observables do not depend on the choice of gauge condition.

The form of the WZ action depends on the regularization, the difference
being a trivial 1-cocycle. Although the difference is a trivial 1-cocycle
it may change drastically the physical content of the theory, as we have
already seen in the chiral Schwinger model. From the point of view of
physical applications it would be of interest to analyze the freedom in
choosing the modified action related  to nonuniqueness of the WZ action.
Below we shall show that in the case of $ SU(N) $ chiral gauge models one
can choose a regularization preserving the gauge invariance with respect
to $SO(N)$ subgroup and calculate the corresponding WZ action, the anomaly
and the 2-cocycle appearing in the commutator of the Gauss law \cite{12}.
We consider in details the canonical quantization of the $SU(3)$ model. In
this case the symplectic form of the WZ action is degenerate and the
quantization procedure which has been used for nondegenerate case by
Faddeev and Shatashvili \cite{3} should be modified. Due to degeneracy
of the symplectic form secondary constraints appear which together with
the primary constraints form a set of secondary class ones. We construct
a path integral representation for the generating functional in the
temporal gauge and prove the gauge invariance of physical observables
\cite{13}.

The paper is organised as follows. In the second part we derive the
expression for the $SO(N)$ invariant WZ action. We present also an
alternative expression for the WZ action in terms of the chiral fields
with values in the coset space $SU(N)/SO(N)$. Using this action we
calculate the anomaly. In the third section we calculate the infinitesimal
2-cocycle appearing in the anomalous constraints commutator and establish
that it vanishes on the $so(N)$ subalgebra. The fourth section is devoted
to the canonical quantization of the WZ actions with degenerate symplectic
forms. We consider as the simplest example the two dimensional chiral
$SU(2)$ gauge model. In the two dimensional case there is a family of WZ
actions parametrized by one parameter $a$ and the choice of $a=0$
corresponds to the WZ action with the degenerate symplectic form. As was
mentioned by Shatashvili \cite{14} this case differs from the others and
requires a special analysis.In the fifth section we discuss the four
dimensional model with the $SO(3)$ invariant WZ action. We point out  a
parametrization of the coset space $SU(N)/SO(N)$ reducing the WZ action to
a pure four dimensional one and then perform canonical quantization. As
was expected we find new physical degrees of freedom but contrary to the
standard WZ action we get not four but two new degrees of freedom.

\section{ The SO(N) invariant WZ action}

\setcounter{equation}{0}

We consider the model described by the classical action
\be
S=\int d^4x\,[-\frac14(F_{\mu \nu }^a)^2+\ls],
\label{2.1d}
\ee
Where $A_\mu $  is an $SU(N)$ Yang-Mills field and $\psi \equiv \frac 12
(1+\gamma_5)\psi$            is a chiral
fermion in the fundamental representation of the $SU(N)$ generated by the
antihermitian matrices $\lambda^a $
\be
\tr{\la\,\lb}=-\frac{1}{2}\delta_{ab};\quad
[\la,\lb]=f^{abc}\, \lambda^{c},
\label{2.2d}
\ee

The action (\ref{2.1d}) is invariant with respect to the gauge
transfomations,
\bea
&&\am\,\rightarrow  \,\am^{g}={g}^{-1}\am
{g}+{g}^{-1}\partial_{\mu}{g}\nonumber\\
&&\psi\,\rightarrow\,\psi^{g}={g}^{-1}\,\psi,
\qquad {g}\in {SU(N)}.
\label{2.3d}
\eea
Due to the gauge invariance of the action (\ref{2.1d}) in the classical
theory one can impose any admissible gauge condition. However it is
well known that because of the quantum anomaly in quantum theory the
equivalence of different gauges is lost which leads to the inconsistency
of the model.

The existence of the quantum anomaly is related to the absence of a
$SU(N)$ invariant regularization of the action (\ref{2.1d}). However it is
known that $SU(N)$ group has nonanomalous subgroups, the $SO(N)$ subgroup
being a maximal one. It suggests that there exists a regularization
preserving the invariance with respect to the $SO(N)$ subgroup. Indeed,
such a regularization can be described by the following Lagrangian \bea
&&{\cal L}=\ls+\sum_{r=1}^{2K-1}[{i}\prb \gamma ^\mu (\partial _\mu +\am
)\pr -M_{r}\prt C \pr -M_{r}\prb C \prbt] \nonumber \\
&&\quad+{i}\sum_{r=1}^{2K}[(-1)^r \phrb \gamma ^\mu (\partial  _\mu + \am
)\phr-\sum_{r,s=1}^{2K}(M_{rs}\phrt C\phi_s - M_{rs}\phrb C\bar{\phi}_s^T
)] \label{4a} \eea Here $\pr $ are the anticommuting Pauli--Villars
spinors  and $\phr $ are the commuting ones. $M_{rs}$ is an antisymmetric
matrix. The standard Pauli--Villars conditions are assumed. The matrix $C$
is the charge conjugation matrix. The only  terms, which  are not
invariant under the gauge transformation (\ref{2.3d}) of all fields, are
the mass terms for the Pauli--Villars fields.  The mass terms transform
as follows \be M_{r}\prb C\prbt \rightarrow \,M_{r}\prb Cgg^{T} \prbt .
\label{5a}
\ee
One sees that for $g\in SO(N)$, $gg^{T}=1$ this mass term is invariant, and
therefore the regularization preserves the $SO(N)$ gauge invariance.


It follows that the anomaly calculated with the
help of this regularization vanishes on the $so(N)$ subalgebra. The
standard anomaly \cite{b} does not possess this property. Anomaly can be
defined as a gauge variation of the functional called the WZ action given
by the following equation,
\be
\hbox{e}^{{i}\a_1 (A,g)}=\frac{\det (\gamma^\mu(\partial_\mu
+\am^g))}{\det (\gamma^\mu(\partial_\mu +\am))} .
\label{6a}
\ee

The value of this determinants depend on the particular regularization
used. Below we shall calculate the WZ action corresponding to the
regularization (\ref{4a}) assuming that necessary counterterms are
introduced. It follows directly from eq.(\ref{6a}) that the WZ action
satisfies the condition
\be
\a_1 (A,g_1)+\a_1 (A^{g_1},g_2)=\a_1 (A,g_1 g_2)\qquad (mod\, 2\pi).
\label{7a}
\ee
This equation is a definition of a 1-cocycle on the gauge group (see
Appendix A). Different regularizations  lead to the WZ actions which
differ by a trivial 1-cocycle. We call cocycle to be trivial if it can be
presented as a finite gauge variation of a local functional $\alpha_0(A)$
(0-cochain),
\begin{equation}
\alpha_1^{triv} (A, g) = \alpha_0(A^g) - \alpha_0(A)
\label{2.8}
\end{equation}
It is worthwile to note that in anomalous theories the WZ action is not a
trivial 1-cocycle.

In the case of regularization (\ref{4a}) the WZ action $\aor $
possesses the $SO(N)$ invariance \be \aor (A,gh)=\aor (A,g), \label{8a}
\ee
where $h\in SO(N)$.

Let us stress that in eq.(\ref{8a}) the field $A$ is not transformed.
Eq.(\ref{8a}) is a direct consequence of the invariance of the gauge
transformed mass term (\ref{5a}) under the transformation
\be
g\rightarrow
gh,\qquad h\in SO(N).
\label{9a}
\ee
Eq.(\ref{8a}) expresses the hidden
symmetry of the Wess--Zumino action in our case. Hidden symmetries of this
type in connection with models on homogeneous spaces were  discussed in
 refs.\cite{CJ,EF,Wu}. It follows from eqs.(\ref{7a},\ref{8a}) that the
Wess--Zumino action vanishes if the chiral field $g$ belongs to the
orthogonal subgroup $SO(N)$
\be
\aor (A,h)=0,\qquad h\in SO(N)
\label{10a}
\ee
The geometric origin of the existence of such Wess--Zumino action is the
triviality of the cohomology group $H^5(SO(N))$.

To calculate the $SO(N)$ invariant WZ action we shall use the fact  that
as was discussed above any WZ action can be presented in the form
\be
\aor (A,g)=\alpha_1 (A,g) + \alpha_0 (A^g) -\alpha_0 (A).
\label{11a}
\ee
Here $\alpha_1 (A,g)$ is the "standard" Wess--Zumino action
\bea
\alpha_1 (A,g)&=&\int d^4x\,[d^{-1}\kappa (g)-\frac {i}{48\pi^2}
\epsilon^{\mu \nu \lambda \sigma }\tr[(\am \partial_\nu \al
\,+\,\partial_\mu \an \al \,+\,\am \an \al )g_\sigma \,-\nonumber \\
&&-\,\frac 12\am g_\nu \al g_\sigma \,-\,\am g_\nu g_\lambda
g_\sigma]]
\label{12a}
\eea
and we use the notations
\be
\int d^4xd^{-1}\kappa (g) \equiv -\frac{i}{240\pi^2}\int_{M_5} d^5x \,
\epsilon^{pqrst} \tr{(g_p
g_q g_r g_s g_t)}
\label{13a}
\ee
\be
g_\mu=\partial_\mu gg^{-1}.
\label{14a}
\ee
In eq.(\ref{13a}) the  integration goes  over a  five-dimensional manifold
whose boundary is the usual  four-dimensional space.

The functional $\alpha_0 (A^g)- \alpha_0 (A)$ is a trivial local  1-cocycle
which can be determined  from eq.(\ref{10a}). The explicit  form of
$\alpha_1 (A,g)$ (eq.(\ref{12a})) dictates the following  ansatz for
$\alpha_0 (A)$:
\bea
\alpha_0 (A)&=&-\frac {i}{48\pi^2}\int  d^4x\,
\epsilon^{\mu \nu \lambda \sigma } \tr(a_1 \am \an \al \ast+a_2 \am  \ant
\al \ast +\nonumber\\ &&+a_3 \am \an \alt \ast + b_1 \partial_\mu \an \al
\ast + b_2 \partial_\mu \an \alt \as + b_3 \partial_\mu \an \alt \ast )
\label{15a}
\eea
where $\amt$ is a transposed  matrix $\am $. We choose this ansatz because
it is the most general local functional which is metric independent and is
invariant under global $SO(N)$ transformations.

Let us stress that to satisfy eq.(\ref{10a}) it is necessary to introduce
the terms depending not only on $\am$ but also on $\amt$. Eq.(\ref{10a})
determines uniquely the coefficients $a_i,b_i$. As a result:
\bea
\alpha_0 (A)&=&-\frac {i}{48\pi^2}\int  d^4x\, \epsilon^{\mu \nu \lambda
\sigma } \tr( \am \an \al \ast-\frac 14 \am  \ant  \al \ast +\nonumber\\
&& + \partial_\mu \an
\al \ast + \am \partial_\nu \al \ast )
\label{16a}
\eea
Obviously one can add  also any trivial local $SO(N)$ invariant 1-cocycle.
The corresponding infinitesimal 1-cocycle (anomaly) is calculated in a
standard way
\be
\int d^4x\,\epsilon^a(x)\, {\cal A}^{a}_{ort}(A)=\aor (A^h,h^{-1}g)-\aor(A,g)
\label{17a}
\ee
where $h=1+\epsilon^a \lambda^a$.

It looks as follows
\bea
&&{\cal A}^{a}_{ort}(A)=\frac {i}{48\pi^2}\epsilon^{\mu \nu \lambda \sigma }
\tr[( \lambda^a+\lambda^{a,T})(\partial_\mu (\an \partial_\lambda
\as+\nonumber \\
&&\quad+\partial_\nu \al \as+\an \al \as -\an \alt \as -\frac 12  \ant
\partial_\lambda\as-\frac 12\partial_\nu  \al \ast)-\nonumber\\
&&\quad-\partial_\mu \an \al \ast-\am \partial_\nu \al \ast-\amt \partial_\nu
\al
\as-\amt \an\partial_\lambda \as-\nonumber \\
&&\quad-\am \an \al \ast+\frac 12 \am \ant \al \ast+\frac 12 \amt \an \alt
\as)]
\label{18a}
\eea
One sees that on the subgroup $SO(N)$ $(\lambda^a=-\lambda^{a,T})$  this
anomaly vanishes. The Wess--Zumino consistensy condition \cite{9}
(infinitesimal verstion of (\ref{7a})) is obviously satisfied because our
anomaly differs from the standard one by the trivial 1-cocycle.

The additional $SO(N)$ invariance of the Wess--Zumino action $\aor(A,g)$
means that it depends in fact not on all the elements   of  $SU(N)$ but only
on the elements of the homogeneous space $SU(N)/SO(N)$. One can introduce
coordinates on this homogeneous space and express the Wess--Zumino action
in terms of these coordinates.

The natural coordinates are symmetric  and unitary matrices
\be
s=gg^T
\label{19a}
\ee
This choice is suggested by the form of the mass term in the regularized
Lagrangian (\ref{14a}). As follows from eq.(\ref{5a}) after the gauge
transformation it  depends only on the combination  $gg^T$. The gauge group
transforms the coordinates $s$ in the following manner
\be
s\rightarrow  g^{-1}sg^{-1,T}.
\ee
 In terms of these coordinates the Wess--Zumino action looks
as follows:
\bea
\aor&=&\int d^4x\,[\frac 12d^{-1}\kappa (s)-\frac {i}{48\pi^2}
\epsilon^{\mu \nu \lambda \sigma }\tr[(\partial_\mu \an \al
+ \am \partial_\nu \al +\am \an \al-\nonumber \\
& &-\frac 12 \partial_\mu
\an s\alt s^{-1}- \frac 12 s \amt s^{-1} \partial_\nu \al- \am s\ant
s^{-1}\al)s_\sigma-\nonumber \\
& &-\frac 12 \am s_\nu \al s_\sigma +\frac 12 (s\amt s^{-1}\an-\am s\ant
s^{-1})s_\lambda s_\sigma - \am s_\nu s_\lambda s_\sigma \nonumber \\
& &+\partial_\mu \an \al s\ast s^{-1}+ \am \partial_\nu \al s\ast s^{-1} +
\am \an \al s\ast s^{-1}-\nonumber \\
& &-\frac 14 \am s \ant s^{-1}\al s\ast s^{-1}-\alpha_0 (A)]]
\label{cocycle}
\eea
where $s_\mu=\partial_\mu ss^{-1}$.

The derivation is straightforward but some comments are  in order. Using the
equality
\be
g_\mu^T=s^{-1}(s_\mu-g_\mu)s
\label{21a}
\ee
we express $g_\mu^T$ in terms of $g_\mu$ and $s_\mu$ and then comparing  the
terms of a given order in $\am$ and applying again eq.(\ref{21a}) we find
the expression (\ref{cocycle}). This action may be used for the
construction of the symplectic form defining the integration measure in
the path integral.  It is worthwhile to emphasize that contrary to  the
standard case the action (\ref{cocycle}) depends not only on the chiral
current $\partial_\mu ss^{-1}$, belonging  to the Lie algebra of the
group, but also on the coordinates of the homogeneous space $SU(N)/SO(N)$.
It  may be of importance for analyzing possible stationary points of the
effective action.

\section {Anomalous constraints commutator}

\setcounter{equation}{0}

In this section  we shall calculate the 2-cocycle
associated to the Wess--Zumino action (\ref{11a}). This 2-cocycle appears
as the Schwinger term  in the constraints commutator and can be calculated
either by direct summation of the Feynman diagrams \cite{KSS,Jo} or by
using the path integral representation for the commutator \cite{AMF}. We
use the second approach.  According to the Bjorken-Johnson-Low (BJL)
formula the matrix element of the equal time commutator may be expressed
in terms of the expectation value of $T$-product as  follows:  \be
\lim_{q_0 \rightarrow\infty} q_0 \int dt'\,\hbox{e}^{iq_0(t'-t) }\langle
\tilde{\varphi}
|\,TA(x,t')B(y,t)\,|\varphi  \rangle =i\langle \tilde{\varphi}
|\,[A(x,t),B(y,t)]\,|\varphi  \rangle
\label{22a}
\ee

For the expectation value of $T$--product one can write the representation
in terms of the path integral
\be
\langle \tilde{\varphi}
|\,TA(x,t')B(y,t)\,|\varphi  \rangle =\int d\mu\,\hbox{e}^{iS}A(x,t')B(y,t)
\label{23a}
\ee
Here it is understood that the integration goes over the fields satisfying the
boundary conditions corresponding to the initial and final states $|\varphi
\rangle$ and $\langle \tilde{\varphi} |$. Following the approach of
\cite{AMF} we can consider the chiral $SU(N)$ Yang--Mills model in the
Hamiltonian gauge $A_0=0$. In this gauge the $S$--matrix element can be
written as the path integral
\be
\langle \alpha |\beta \rangle=\int d\mu \,\delta  (A_0) \hbox{e}^{iS},
\label{24a}
\ee
where in the first order formalism
\bea
S&=&\int d^4x\,[E_i^a\dot{A}_i^a-\frac 12 {(E_i^a})^2-\frac 14{(F_{ij}^a)}^2+
A_0^a G^a+\nonumber \\
&&+i\bar{\psi} \gamma_0 \partial_0\psi-i\bar{\psi} \gamma_i
(\partial_i-A_i)\psi ]
\label{26a}
\eea

In the nonanomalous  case the constraints $G^a = (\nabla_i E_i)^a $ form a
Lie algebra \be [G^a({\bf x}),G^b({\bf y})]=if^{abc}G^c({\bf y})\delta
({\bf x}-{\bf y}) \label{27a} \ee

However as  was shown in refs. \cite{AMF,KSS,Jo} in the anomalous theory this
relation is violated and the Schwinger term arises.

To calculate this Schwinger term we make the gauge transformation of the
variables in the integral (\ref{24a}). The transformed integral   may be
written in the form
\be
\langle \tilde{\varphi} |\varphi \rangle=\int d\mu \, \delta  (A_0)
\hbox{e}^{iS}\exp \{-i\int  d^4x\,g_0^a G^a(x)+i\aor
(A,g)\bigm|_{A_0=-g_0}\}
\label{28a}
\ee
Here the 1-cocycle arises due to the noninvariance of the regularization in
accordance  with eq.(\ref{6a}).

Using the  representation for  the chiral field $g$: $g=\hbox{e}^u$ and taking
into  account  that the integral (\ref{28a}) does not  depend on $g$ we can
put equal to zero variation of this integral over  $u$. To the second order
in $u$ one has:
\bea
&&\frac 12 \int d^4x d^4y \,\langle \tilde{\varphi} |T\widetilde{G}^a(x)
\widetilde{G}^b(y)|\varphi  \rangle \partial_0 u^a(x) \partial_0
u^b(y)+\nonumber \\
&&+\frac i2 \int d^4x \,f^{abc}u^a(x) \partial_0 u^b(y)\langle \tilde{\varphi}
|T\widetilde{G}^a(x) |\varphi  \rangle+\nonumber \\
&&+\frac {1}{48\pi^2} \int d^4x\,\langle\tilde{\varphi}| \tr(\epsilon^{ijk}
\partial_i A_j \{\partial_k u(x),  \partial_0 u(x)\}) |\varphi
\rangle+ ...  =0 .
\label{29a}
\eea
Here we introduced the notation
\bea
\widetilde{G}^a(x)&=&G^a(x)-\frac {i}{48\pi^2}\epsilon_{ijk}\tr[(\lambda^a  +
\lambda^{a,T})(A_i \partial_j A_k+\partial_i A_jA_k+A_iA_jA_k-A_i A_j^T
A_k)\nonumber   \\
& &\quad-\lambda^a \{\partial_i A_j,A_k^T\}]
\label{30a}
\eea
And ...  denotes the terms which do not contribute in the BJL limit.

In the process of derivation of eqs.(\ref{29a}, \ref{30a}) we used the
explicit form of $\aor $ (\ref{11a}, \ref{16a}) up to the second order in
$u$, \bea &&\aor(-g_0,A_i,g_i)=\frac {i}{48\pi^2}\tr \int dt\, \int \{
(AdA+dAA+A^3-AA^TA+(T)-\nonumber \\ &&-\{ dA,A^T\} (\partial_0 u+\frac 12
[u,\partial_0 u])+ +(2dA+A^2)(du\partial_0 u+\partial_0 u du)+ \nonumber
\\ &&-2A\partial_0uAdu-\partial_0 A [A,du+\frac 12 [u,du]]\} + +\{
(dg_0+[A,g_0])(g[A^g,A^{T,g}]g^{-1}- \nonumber \\ && \quad
-[A,A^T]-\partial_0A(g[A^g,A^{T,g}]g^{-1}-[A,A^T])\}.  \label{B.4} \eea
where $(T)$ stands for transposition of the first four terms. The
coefficients of $\partial_0 A $ can be combined with $E$ leading to the
shift of $E_i^a $, \be E_i^a \rightarrow E_i^a+\frac
{i}{48\pi^2}\epsilon_{ijk}\tr
\lambda^a(\{A_j,g_k\}+g\{A_j^g,A^{T,g}_k\}g^{-1}-\{A_j,A_k^T\})
\label{31a}
\ee
Combining the coefficients of $g_0 $ with $G^a ({\bf x})$ we get
\bea
&&\int d\mu \, \exp[i\int d^4x\, (E_i^a
\dot{A}_i^a-\widetilde{H}(E_i,A_i,\partial_i A_j )-g_0^a\widetilde{G}^a(x))+
\nonumber \\
&&\quad +\frac{i}{48\pi^2}\int dt\int \, dA(du\partial_0 u+\partial_0 u du)]
\label{B.6}
\eea
Eq. (\ref{29a}) follows directly from eq. (\ref{B.6}).

To get the expression for the commutator of  $\widetilde{G}$ we apply to
eq.(\ref{29a}) the operator:
\be
\lim_{(p_0-q_0)\rightarrow\infty}\frac {p_0-q_0}{p_0q_0}\int
dx_0dy_0\,\hbox{e}^{ip_0x_0+iq_0y_0} \frac {\delta}{\delta  u^a(x)}\frac
{\delta}{\delta  u^b(y)}
\label{32a}
\ee
Taking the limit we  get  the result
\be
[\widetilde{G}^a({\bf x}),\widetilde{G}^b({\bf y})]=if^{abc}\widetilde{G}^c
({\bf y})\delta ({\bf x}-{\bf y})
-\frac {1}{24\pi^2} \epsilon_{ijk}\tr (\partial_i A_j\{\lambda^a
,\lambda^b\})\partial_k^x \delta ({\bf x}-{\bf y}).
\label{33a}
\ee

Let us note that the commutator  of $\widetilde{G}$ (\ref{33a}) coincides
with the analogous commutator obtained  in ref.\cite{AMF} with the different
Wess--Zumino  action. However the definition  of  $\widetilde{G}$ in our case
is different. If one comes  back to the $G$ one gets
\be
[G^a ({\bf x}),G^b ({\bf y})]=if^{abc}G^c({\bf y})\delta
({\bf x}-{\bf y})+a_{2,ort}^{ab} (A;{\bf x},{\bf y})
\label{34a}
\ee
Here $a_{2,ort}^{ab} $  is the ultralocal 2-cocycle
\bea
&&a_{2,ort}^{ab} (A;{\bf x},{\bf y})=-\frac {1}{48\pi^2}\epsilon_{ijk}\tr
([\lambda^a+\lambda^{a,T},\lambda^b+\lambda^{b,T}]\times \nonumber\\
&&\quad\times(A_i\partial_jA_k+
\partial_iA_jA_k
+A_iA_jA_k-A_iA_j^TA_k+A_i^T\partial_jA_k+\partial_iA_j^TA_k)+\nonumber \\
&&\quad+(\lambda^a+\lambda^{a,T})(\partial_iA_j-\partial_iA_j^T-A_iA_j^T-A_i^T
A_j) (\lambda^b+\lambda^{b,T})A_k-\nonumber \\
&&\quad-(\lambda^b+\lambda^{b,T})(\partial_iA_j-\partial_iA_j^T-A_iA_j^T-A_i^T
A_j)
(\lambda^a+\lambda^{a,T})A_k )
\label{35a}
\eea

This cocycle differs from the one obtaned in
ref.\cite{AMF}--\cite{Jo} by trivial 2-cocycle. It vanishes
if at least one of the constraints $G^a$ corresponds to the subgroup
$SO(N)$.  We note that the addition of any trivial 1-cocycle
having topological nature does not change the commutator of modified
constraints $\widetilde{G}$.

Eq.(\ref{34a}) shows that the constraints commutator does not vanish on
the constraint surface and therefore the quantization in the temporal
gauge is inconsistent. To avoid this problem we shall add following the
approach of Faddeev and Shatashvili the WZ action \cite{3} to the classical
action (\ref{2.1d}).
As the gauge variation of the WZ action compensates the anomaly, one can
hope that the quantization of the modified action will lead to a
consistent theory. We continue to use the temporal gauge $A_0=0$ in spite
of the fact that the modified classical action is not gauge invariant. We
shall show that when the quantum corrections are taken into account the
gauge invariance is restored.

The WZ action is the first order action for the chiral fields and to
quantize it one needs to find the symplectic form. If this form is
nondegenerate the quantization is performed in a standard way and has
been done in ref. \cite{3}. However in the case of $SO(3)$ invariant
WZ action considered above the symplectic form is degenerate and the
quantization requires more careful analysis. To illustrate the main ideas
we consider firstly more simple case of two dimensional $SU(2)$ theory
with a degenerate symplectic form.

\section {Quantization of two dimensional SU(2) model}

\setcounter{equation}{0}

The two dimensional chiral $SU(2)$ Yang-Mills theory is described by the
action
\be
S_{YM}=\int d^2x\,(-\frac {1}{4e^{2}} (F_{\mu\nu}^{a})^{2}
+ i\bar{\psi}\gamma^\mu (\partial_\mu+A_{\mu})\psi),
\label{0}
\ee
where $\gamma^0=\sigma^1$, $\gamma^1=\sigma^2 $ and $ \psi \equiv
\frac{(1+\gamma_5)}{2} \psi $, $ \gamma_5 \equiv i\sigma^1 \sigma^2
=-\sigma^3 $. The algebra of two dimensional $\gamma$-matrices allows to
rewrite this action in the form
\be
S_{YM}=\int d^2x\,(-\frac {1}{4e^{2}}
(F_{\mu\nu}^{a})^{2} + i\psi^{+}(\partial_{+}+A_{+})\psi),
\label{1}
\ee
where
$\partial_{+}=\partial_{0}+\partial_{1}; \qquad A_{+}=A_{0}+A_{1}$.

On the classical level this action possesses the usual gauge invariance,
however as is well-known quantum corrections violate this invariance. To
restore the gauge invariance one can following Faddeev and Shatashvili \cite{3}
add to the action (\ref{0}) the corresponding \wz action, which in our
case looks as follows:  \bea S_{WZ} & = & -\frac{1}{12\pi} \int_{M^{+}}
d^3x\,\e^{ijk}\tr g_{i}g_{j}g_{k}+ \frac{1}{4\pi}\int d^2x\, \e^{ \mu \nu}
\tr ( g_{\mu} A_{\nu}+ \nonumber \\ & + & \frac{a}{2} (\am+g_\mu )^2 -
\frac{a}{2} (\am )^2 ) \label{2} \eea Here $g_{i}=\partial_{i}
gg^{-1},\quad g \in SU(2),\quad\e^{ijk} $ and $\e^{\mu \nu }$ are
antisymmetric tensors, $M^{+}$ is a three-dimensional manifold whose
boundary is the usual two-dimensional space and $a$ is an arbitrary
parameter depending on the regularization used to calculate the WZ action.

If $a$ is different from zero the action is nondegenerate and this case
was considered  by Shatashvili \cite{14}. In particular when $a=-1$ the
model is exactly soluble \cite{16}. The case $a=0$ is exceptional.In this
case the WZ action does not depend on the space-time metrics and its
symplectic form is degenerate (this is true for any gauge group). Below we
carry out the Hamiltonian analysis of this case.

Following the strategy discussed in the Introduction we impose some gauge
condition (here we shall use the temporal gauge $A_0=0$), apply the
canonical formalism and construct the path integral
representation for the generating functional. Then we prove the
gauge invariance of the integration measure justifying thus the
possibility of imposing gauge condition before the quantization.

The first problem in applying the canonical quantization is the three
dimensional term in the \wz action (\ref{2}). It is known that this term
depends only on the values of the chiral field $g$ on the two-dimensional
boundary (more exactly by $mod \, 2\pi$) and therefore one  can choose
such a parametrization of the field $g$ in which this term can be written
exlicitely as a two-dimensional one.  We use the parametrization of the
$SU(2)$ group by the fields $\phi^{A}$, satisfying the following
condition:
\be
\tr (g_{A}g_{B}g_{C})=6\pi\e_{ABC}, \label{3} \ee where
$g_{A}=\frac {\partial g}{\partial \phi^{A}}g^{-1}=\partial_{A}gg^{-1}$ is
a right-invariant vector field on the $SU(2)$ group.

In terms of the fields $\phi^{A}$ any right-invariant current $g_{i}$ can be
expressed by the following formula:
\be
g_{i}=\frac {\partial g}{\partial x^{i}}g^{-1}=g_{A}\partial_{i}\phi^{A}
\label{4}
\ee
Due to the condition (\ref{3}) the Haar measure $dgg^{-1}$ on the $SU(2)$
group is proportional to $d\phi^{1}d\phi^{2}d\phi^{3}$.

Using the parametrization by the fields $\phi^{A}$ and imposing the light-cone
gauge one can rewrite the sum of (\ref{1}) and (\ref{2}) as follows:
\bea
S&=&\int d^2x\,(\frac {1}{2e^{2}} (\partial_{0}A^{a})^{2} +
\frac {1}{2}\e_{ABC}\e^{\mu\nu}\phi^{A}\partial_{\mu}\phi^{B}
\partial_{\nu}\phi^{C}
+\frac {1}{4\pi}\tr (g_{A}A)\partial_{0}\phi^{A} + \nonumber \\
&+& i\psi^{+}\partial_{0}\psi + i\psi^{+}\partial_{1}\psi + i\psi^{+}
A \psi )
\label{5}
\eea
Here $A=A_{1}$.

Introducing the canonically-conjugated momenta for the fields $A$ and
$\phi^{A}$
one can present the action (\ref{5}) in an equivalent form:
\bea
S&=&\int d^2x\,(E_{a}\partial_{0}A^{a}+p_{A}\partial_{0}\phi^{A}
-\frac 12 (E_{a})^{2} +
\nonumber\\
&+&\lambda^{A}(p_{A}+\e_{ABC}\phi^{B}\partial_{1}\phi^{C}-
\frac {1}{4\pi}\tr (g_{A}A) + \nonumber \\
&+& i\psi^{+}\partial_{0}\psi + i\psi^{+}\partial_{1}\psi + i\psi^{+}
A \psi )
\label{6}
\eea
{}From (\ref{6}) one can conclude that
\be
H=\frac 12 (E_{a})^{2}-
i\psi^{+}\partial_{1}\psi - i\psi^{+}A \psi
\label{7}
\ee
is the Hamiltonian and
\be
C_{A}=p_{A}+\e_{ABC}\phi^{B}\partial_{1}\phi^{C}-\frac {1}{4\pi}\tr (g_{A}A)
\label{8}
\ee
are the primary constraints of the model.

The next step in the canonical quantization is the calculation of  secondary
constraints. The simplest way seems to be to find all null-vectors of the
matrix of the Poisson brackets of the primary constraints. Then for every
null-vector
$e_{\alpha}^{a}$ one can form a linear combination of the primary constraints
$C_{\alpha}=C_{A}e_{\alpha}^{A}$, which commutes with all primary constraints
on the constraints surface. The secondary constraints are then given by the
the Poisson brackets of $C_{\alpha}$ and the Hamiltonian $H$.

In our case the matrix of the Poisson brackets the primary constraints is equal
to:
\bea
\Omega_{AB} (x^{1},y^{1})&=&\{ C_{A}(x^{1}),C_{B}(y^{1})\} =
\Omega_{AB} (x^{1})\delta(x^{1}-y^{1})\nonumber\\ &=&
\frac {1}{4\pi}\tr ([g_{A},g_{B}](g_{1}(x^{1})+A(x^{1})))\delta(x^{1}-y^{1})
\label{9}
\eea
This matrix is ultralocal and in fact coincides with the symplectic form for
the
\wz action. There is only one null-vector of $\Omega_{AB} $ (in every space
point)
equal to
\be
e^{A}(x^{1})=\frac {1}{4\pi}\e^{ABC}\tr ([g_{B},g_{C}](g_{1}(x^{1})+A(x^{1}))=
\e^{ABC}\Omega_{BC} (x^{1})
\label{10}
\ee
Calculating the Poisson bracket of the constraint
 $\widetilde{C}(x^{1})= C_{A}(x^{1})e^{A}(x^{1})$
and $H$ one gets up to the primary constraints the secondary constraint:
\be
C(x^{1})=4\pi\{ H, \widetilde{C}(x^{1})\} =\tr (E(x^{1})(g_{1}(x^{1})+A(x^{1}))
\label{11}
\ee
 In this equation we omited the term proportional to $C_A$.
The primary constraints $C_{A}(x^{1})$ and the secondary constraint $C(x^{1})$
form a set of second-class constraints and the matrix of the Poisson brackets
of
the constraints is equal to:
\be
M(x^{1},y^{1})=\left( \begin{array}{cc} \Omega_{AB} (x^{1},y^{1}) &
v_{A}(x^{1},y^{1}) \\
-v_{B}(y^{1},x^{1}) & 0 \end{array} \right)
\label{12}
\ee
where
\bea
v_{A}(x^{1},y^{1})&=&\{ C_{A}(x^{1}),C(y^{1})\} \nonumber\\ &=&
\tr (g_{A}(\partial_{1} E-[g_{1},E] - \frac {1}{4\pi} (g_{1}+A)))
\delta(x^{1}-y^{1})\nonumber\\ &\quad &-\tr
(g_{A}E(x^{1}))\partial_{1}^{y}\delta(x^{1}-y^{1})
\label{13}
\eea
It is not difficult to show that the determinant of the matrix $M$ is equal
to
\be
\det M=(\det \e^{ABC}\Omega_{AB}v_{C})^{2}=(\det e^{A}v_{A})^{2}
\label{14}
\ee
and
\be
e^{A}v_{A}(x^{1},y^{1})=
\tr ((g_{1}+A)(\nabla_{1}E-\frac {1}{4\pi}
(g_{1}+A)))\delta(x^{1}-y^{1})
\label{15}
\ee
up to the secondary constraint $C(x^{1})$.
Now one can write the expression for the generating functional of the
model:
\bea
Z&=&\int \,DADED\phi DpD\psi (\det M)^{\frac 12}\delta(C)\delta(C_{A})
\nonumber\\
&\quad &\hbox{exp}\{ i\int d^2x\,
(E_{a}\partial_{0}A^{a}+p_{A}\partial_{0}\phi^{A}
-\frac 12 (E_{a})^{2} + \nonumber \\
&\quad& i\psi^{+}\partial_{0}\psi + i\psi^{+}\partial_{1}\psi + i\psi^{+}
A \psi )\}
\label{16}
\eea
Integrating over $p_{A}$ and introducing the
integration over $A_{0}$ in the path integral one gets \be Z=\int
\,DA_{\mu}DED\phi D\psi (\det M)^{\frac 12}\delta(C)\delta(A_{0})
\hbox{exp}\{ i(S_{YM}+S_{WZ}) \}
\label{17}
\ee
It is obvious from eqs.(\ref{11}, \ref{15}) that the integration measure
in eq.(\ref{17}) is gauge-invariant  apart from  the gauge-fixing
condition and the fermion measure. Therefore one can easily show that
the modified Gauss-law constraints form the $SU(2)$ gauge algebra:  \be
 [G_{a}(x^{1}),G_{b}(y^{1})]=i\e_{abc} G_{c}(x^{1}) \delta(x^{1}-y^{1})
\label{20}
\ee
where
\be
G(x^{1})=\nabla_{1}E(x^{1})-\frac {1}{4\pi}g_{1}(x^{1})+j_{0}(x^{1})
\label{21}
\ee
Indeed in our case the gauge variation of the WZ action exactly
compensates anomaly arising due to noninvariance of the fermionic measure
and all other factors are gauge invariant. Therefore one can repeat all the
arguments given in the proceeding section to show that the Gauss law has
 the form (\ref{20}) .

Due to this fact one can select the physical subspace
imposing the condition $G_{a}|\Psi >=0$ on the state vectors.  The number
of the physical degrees of freedom can be now easily calculated.  All
vector fields are unphysical due to the Gauss-law constraints and there is
only one physical degree of freedom for three chiral fields $\phi_{a}$ due
to the four second-class constraints.

This result is in accordance with the intuitive expectations. In the
classical theory we had the system with three first class constraints
eliminating all the bosonic degrees of freedom. Due to the quantum
anomaly the first class constraints transform to the second class ones
and one secondary constraint arises. Thus we have a system of four second
class constraints eliminating two degrees of freedom. One degree survives
as a physical excitation.

\section{Quantization of the four dimensional SU(3) model}

\setcounter{equation}{0}

Now we are ready to quantize the four dimensional chiral $SU(3)$
 Yang-Mills theory. The complete action of the model is described by the
sum of the Yang-Mills action and the $SO(3)$ invariant WZ action,
\be
S=\int d^4x\,(-\frac {1}{4e^{2}} (F_{\mu\nu}^{a})^{2}
+ \ls )+\aor (A,s)
\label{22}
\ee
 where $\aor $ is given by (\ref{cocycle}).

To apply the canonical formalism to the model one needs, as was mentioned
in the Introduction, to reduce the five-dimensional term in the \wz action
to a four-dimensional one. To do it one can use the fact that any symmetric
unitary matrix can be represented in the following form:
\be
s=\omega D\omega^{T}
\label{26}
\ee
where $\omega$ is an orthogonal matrix $\omega\omega^{T}=1$ and $D$ is a
diagonal unitary matrix.

Using this representation and and the 1-cocycle condition
\be
\aor (A^{h},h^{-1}sh^{-1,T})=\aor (A,s)-\aor (A,hh^{T})\qquad (mod\, 2\pi)
\label{25}
\ee
one can show the
validity of the following equation:
\be
\aor (A,\omega D\omega^{T})=\aor
(A^{\omega}, D)
\label{27}
\ee
The five-dimensional term is equal to zero
for any diagonal matrix and therefore the parametrization (\ref{26})
solves the problem of reducing the \wz action to a four-dimensional form.

Let us now represent the matrix $D$ in the form $D=\hbox{e}
^{u^{\a}T_{\a}}$, where matrices $T_{\a}$ belong to the Cartan subalgebra
of the  $su(N)$ algebra, and use an arbitrary parametrization of the
$SO(N)$ group by fields $\phi^{A}$. Then introducing the
canonically conjugated momenta for the fields $A_{i}$, $\phi^{A}$ and
$u^{\a}$ and imposing the temporal gauge $A_{0}=0$ one can rewrite the
action (\ref{22}) as follows:
\bea
S&=&\int d^4 x (\Pi^i_a \p_0 A_i^a + p_A
\p_0 \phi^A + \pi_\alpha \partial_0 u^\alpha - \frac{1}{2} (\Pi^i_a-\Delta
E^i_a)^2- \nonumber \\ && \frac{1}{4} (F_{ij}^a)^2 + \lambda^\alpha
C_\alpha+ \lambda^A C_A + {\cal L}_\psi )
\label{28}
\eea
where
\bea
\Delta
E^i_a&=&-\frac{i}{48\pi^2}\e^{ijk} \tr (T_a \omega (\{A_j^\omega, u_k
\}-\frac{1}{2} \{DA_j^{\omega,T}D^{-1},u_k \}+ \nonumber \\ && +
\{A_j^\omega, DA_k^{\omega, T} D^{-1} \} - \{A_j^\omega, A_k^{\omega, T}
\})\omega^{-1}) \nonumber \\ A_i^\omega&=& \omega^{-1}A_i \omega +
\omega^{-1} \p \omega;\quad u_i=\p_i u=\p_i DD^{-1}
\label{29}
\eea
and $C_{p}=(C_{\a},C_{a})$ are the
primary constraints of the model
\bea
C_\alpha&=&\pi_\alpha - \frac{i}{48
\pi^2} \e^{ijk} \tr T_\alpha (\{\p_i A_j^\omega ,A_k^\omega \} +
A_i^\omega A_j^\omega A_k^\omega - \nonumber \\ &&-\frac{1}{2} \{\p_i
A_j^\omega, DA_k^{\omega, T} D^{-1} \}- A_i^\omega DA_i^{\omega, T} D^{-1}
A_k^\omega -A_i^\omega u_j A_k^\omega
\label{30}
\eea
\bea
C_A&=&p_A+ \frac{i}{48 \pi^2} \e^{ijk} \tr \omega_A ( \{\p_i A_j^\omega,
u_k \} - \frac{1}{2} D\{\p_i A_j^{\omega, T} , u_k \} D^{-1}+
\nonumber \\
&+&A_i^\omega u_j A_k^\omega -D^{-1} A_i^\omega u_j A_k^\omega D -
DA_i^{\omega, T}D^{-1} A_j^\omega u_k -u_k A_i^\omega DA_j^{\omega ,T}
D^{-1}+
\nonumber \\
&+& \frac{1}{2} [A_i^\omega, \{DA_j^{\omega, T} D^{-1} , u_k \} ] - u_i
A_j^\omega u_k+ \{ \p_i A_j^\omega , DA_k^{\omega , T}D^{-1} - A_k^{\omega
, T} \} +
\nonumber \\
&+& D^{-1} \{A_k^\omega , \p_i A_j^\omega \} D - \{A_k^\omega , \p_i
A_j^\omega \} +D^{-1} A_i^\omega A_j^\omega A_k^\omega D -  A_i^\omega
A_j^\omega A_k^\omega
\nonumber \\
&-& D^{-1}A_i^\omega DA_j^{\omega, T}D^{-1} A_k^\omega D + A_i^\omega
DA_j^{\omega ,T}D^{-1} A_k^\omega)
\label{31}
\eea
As was mentioned
above, the matrix of the Poisson brackets of the primary constraints
coincides with the symplectic form and is equal to:
\bea
\Omega_{pq}({\bf x,y })&=& \{C_p({\bf x}),C_q({\bf y}) \} =
\Omega_{pq}({\bf x}) \delta({\bf x-y})
\nonumber \\
\Omega_{pq}({\bf x})&=& \frac{i}{96 \pi^2} \e^{ijk} \tr ([s_p , s_q](
\frac{1}{2} \{\widetilde{A}_i , \widetilde{F}_{jk} \}- \widetilde{A}_i
\widetilde{A}_k
\widetilde{A}_k )+
\nonumber \\
&+&s_p( \frac{1}{2} \widetilde{F}_{ij} -  \widetilde{A}_i
\widetilde{A}_j)s_q \widetilde{A}_k
-s_q( \frac{1}{2} \widetilde{F}_{ij} -  \widetilde{A}_i
\widetilde{A}_j)s_p \widetilde{A}_k)
\label{32}
\eea
where
\bea
s_\alpha&=&\frac{\p s}{\p u^\alpha}s^{-1}=\omega \lambda_\alpha
\omega^{-1}; \quad s_A=\frac{\p s}{\p \phi^A}s^{-1}=\omega (
\omega_A-D\omega_A D^{-1})\omega^{-1};
\nonumber \\
&&\omega_A=\omega^{-1}\frac{\p \omega}{\p \phi^A}
\label{33}
\eea
and
\bea
\widetilde{F}_{ij}&=& F_{ij}-sF_{ij}^T s^{-1}; \quad \widetilde{A}_i=
A_i+sA_i^T s^{-1}+ s_i
\label{34}
\eea
In parametrization we use the symplectic matrix is an antisymmetric
$5\times 5$ matrix of the following type
\be
\Omega=\left(
\begin{array}{cc} \{C_\alpha ,C_\beta \} & \{C_\alpha , C_A \}
\\ \{C_B ,C_\beta \} & \{C_A , C_B \} \end{array} \right)
\label{omega}
\ee
where $\{C_\alpha ,C_\beta \} $ is nondegenerate block. For any
antisymmetric matrix with nondegenerate block $A$
\be
\left(
\begin{array}{cc} A & f
\\ -f^T & B  \end{array} \right)
\label{matr}
\ee
the equation for null  vectors
\be
\left(
\begin{array}{c} \alpha_r
\\ \beta_r  \end{array} \right)
\label{vec}
\ee
can be reduced to the following equation for the component $\beta_r$
\be
(B+f^T A^{-1} f)\beta_r = 0
\ee
Substituting to this equation the explicit form of $A$, $B$ and $f$ from
eq. (\ref{omega}) we see that for the case of a general position the only
null  vector is equal to
\bea
e^p({\bf x})&=&\e^{pqrst} \Omega_{qr}({\bf x}) \Omega_{st} ({\bf x})
\label{35}
\eea
As before the secondary constraint is
given by the Poisson bracket  of the constraint $\widetilde{C}({\bf x})=
C_{p}({\bf
x})e^{p}({\bf x})$ and the Hamiltonian $H=\int d^3 x (\frac{1}{2}(\Pi_a^i-
\Delta E_a^i)^2 + \frac{1}{4} (F_{ij}^a)^2)$:
\bea
C({\bf x})&=& \{C_p({\bf x}) e^p({\bf x}), H \} \sim \e^{pqrst}
R_p \Omega_{qr}({\bf x}) \Omega_{st} ({\bf y})
\label{36}
\eea
where
\bea
R_p&=&\e^{ijk} \tr s_p ( \{E_i , F_{jk}- \frac{1}{2} sF_{jk}^T
s^{-1}-\widetilde{A}_j \widetilde{A}_k \} + \widetilde{A}_i E_j
\widetilde{A}_k)
\label{37}
\eea
This formula follows from the equation
\bea
\{C_p({\bf x}), E_i^a({\bf y}) \} & \sim & \epsilon^{ijk} \tr s_p(\{ \la
, F_{jk}-\frac12 s F_{jk}^T s^{-1} - \widetilde{A_j}  \widetilde{A_k}
\}- \nonumber \\
& &  \widetilde{A_j} \la \widetilde{A_k} ) \delta ({\bf x}-{\bf y})
\label{comm}
\eea
As it is shown in the Appendix B the secondary constraint transforms
under the gauge transformation as follows:
\be
C({\bf x}) \rightarrow
\det \left( \frac{ \p \phi^p}{\p \widetilde{\phi^q}} \right) C({\bf x})
\ee
 where $\phi^{p}$ are the coordinates of the point $s$ on the coset
space $SU(3)/SO(3)$ ($u^{\a}$ and $\phi^{A}$ in our case) and $\tilde\phi
^{p}$ are the coordinates of the gauge-transformed point
$g^{-1}sg^{-1,T}$. In other words the function $\tilde\phi(\phi)$ defines
the change of the field $\phi^{p}$ under the gauge transformation.
The five
primary constraints $C_{p}({\bf x})$ and the secondary constraint $C({\bf
x})$ form a set of second-class constraints with the following matrix of
the Poisson brackets of the constraints:
\be M({\bf x},{\bf y})=\left(
\begin{array}{cc} \Omega_{pq} ({\bf x},{\bf y}) & v_{p}({\bf x},{\bf y})
\\ -v_{q}({\bf y},{\bf x}) & v({\bf x},{\bf y}) \end{array} \right)
\label{40}
\ee
where
\be
v_{p}({\bf x},{\bf y})=\{
C_{p}({\bf x}),C({\bf y})\} ,\qquad v({\bf x},{\bf y})=\{ C({\bf
x}),C({\bf y})\}
\label{41}
\ee
The matrix $M({\bf x},{\bf
y})$ has the following  gauge transformation law (see Appendix B):
\bea
M({\bf x},{\bf y})
&\rightarrow & \left( \begin{array}{cc} \frac{\p \phi^p}{\p
\widetilde{\phi}^r}({\bf x}) & 0 \\
0 & det \left( \frac{\p \phi}{\p \widetilde{\phi}}({\bf x})\right)
\end{array} \right) \left( \begin{array}{cc} \Omega_{pq} ({\bf x},{\bf y})
& v_{p}({\bf x},{\bf y}) \\ -v_{q}({\bf y},{\bf x}) & v({\bf x},{\bf y})
\end{array} \right) \times
\nonumber \\
&& \quad
\left( \begin{array}{cc} \frac{\p \phi^q}{\p
\widetilde{\phi}^s}({\bf y}) & 0 \\
0 & det \left( \frac{\p \phi}{\p \widetilde{\phi}}({\bf y}) \right)
\end{array} \right)
\label{42}
\eea
Due to
eq.(\ref{42}) $(\det M)^{\frac 12}$ transforms as follows:  \bea
(detM)^{\frac{1}{2}}& \rightarrow & \left(det \frac{\p \phi}{\p
\widetilde{\phi}}\right)^{2} (detM)^{\frac{1}{2}}
\nonumber \\ && \label{43} \eea Now we can prove
the gauge invariance of the integration measure in the path integral for
the generating functional:
\be Z=\int \,DA_{\mu}DED\phi D\psi D\bar{\psi } (\det
M)^{\frac 12}\delta(C)\delta(A_{0}) \hbox{exp}\{ i(S_{YM}+S_{WZ}) \}
\label{44}
\ee
Taking into account eqs.(\ref{comm}) and (\ref{43}) and the transformation
law  $D\phi \rightarrow det \left( \frac{\partial \phi }{\partial
\widetilde{\phi} } \right) D\phi $ we see that the measure $D\phi (\det
M)^{\frac 12}\delta(C)$ is gauge invariant. Hence in the complete analogy
with the discussion at the end of the preceeding section we justified the
possibility of imposing the gauge condition before the quantization and of
selecting the physical subspace by the Gauss-law constraints.

The number of the physical degrees of freedom can be now easily
calculated. Due to the Gauss-law constraints there are $2\times 8$ vector
degrees of freedom (8 is the dimension of $SU(3)$) and due to the six
second-class constraints there are two bosonic degrees of freedom. Let us
remind that in the case of the standard \wz action one would get four
bosonic degrees of freedom and thus these models differ crucially from each
other in spite of the fact that the difference between these \wz actions
is a local trivial 1-cocycle.  Let us finally note that one could use such
a parametrization of the coset space $SU(3)/SO(3)$ that the invariant
measure is proportional to $d\phi^{1}...d\phi^{5}$. In this
case the secondary constraint and $\det M$ are gauge-invariant.

\section{Discussion}

\setcounter{equation}{0}

In this paper we showed that if one modifies the classical action of
anomalous model by adding the corresponding WZ action the theory can be
consistently quantized and new degrees of freedom appear. The number of
new degrees of freedom depends on the particular choice of the WZ action.
In the case of four dimensional $SU(N)$ gauge models the minimal number of
new degrees of freedom arises if one uses the $SO(N)$ invariant WZ action
described above. At present we have no reliable calculation scheme that
makes difficult more detailed analysis of the model. One could try to
develope some perturbation expansion in terms of vector fields as for $\am
=0$ the WZ action is reparametrization invariant and desribes the exactly
soluble model. Unfortunately the point $\am =0 $ is a singular one and
this procedure fails. In the case of two dimensional model  the
perturbative expansion in the coupling constant can be developed in the
light-cone gauge. However to get really interesting results one needs some
nonperturbative approach which at present is not known.

Finally we mention that the path integral representation for generating
functionals in models considered above can be written in an alternative
form as the path integrals of the exponent of the Lagrangian action. As
was shown in the paper \cite{17} in the path integral for systems
with secondary constraints one can make a canonical transformation
eliminating the secondary constraint and thus to rewrite it as the
integral of the exponent of the Lagrangian action. The price one pays for
it is the appearances of the new local measure in the path integral. In
this approach the symmetry properties  of the action in particular gauge
invariance are manifest but one needs to study the new local measure. In
perturbation theory this measure can be done trivial but in a general case
the problem remains open.

The analysis of the models described above was simplified due to the fact
that in these cases the number of chiral filds was equal to $d+1$ where $d$
is space-time dimension. However it can be generalized to arbitrary
$SU(N)$ chiral model.

{\bf Acknowledgements:} One of the authors (S.F.) would like to thank
Professor J.Wess for kind hospitality and the Alexander von Humboldt
Foundation for the support. This work has been supported in part by
ISF-grant MNB000 and by the Russian Basic Research Fund under
grant number 94-01-00300a.

\appendix
\section {APPENDIX}
\setcounter{equation}{0}

In this Appendix we shall give some information about the cocycles and
show how the $ SO(N) $ invariant WZ action can be obtained in the
framework of descent procedure.

Let us consider a group $ G $ acting on the space $ M $, i.e. $ g:\, m
\rightarrow mg $, $g \in G $, $m \in M$ and real functions $\alpha_0(m)$,
$\alpha_1(m;\,g)$, ..., $\alpha_n(m;\,g_1,\, ...,\, g_n)$, $...$ form the
sequence of functions depending on the point $ m $
and on the ordered set $g_1, ...,g_n $. The operator $ \d $ which acts from
the space with number $ n $ to the space with number $ n+1 $ according to
the following rule
\bea
(\delta \alpha_n)(m;\,g_1,\,...,\,g_{n+1})&=&\alpha_n(mg_1;\,g_2,\,...,
\,g_{n+1})-\alpha_n(m;\,g_1g_2,\,...,\,g_{n+1})+... \nonumber \\
& &\quad+(-1)^n\alpha_n(m;\,g_1,\,g_2,\, ...,\, g_n)
\label{2.20}
\eea
is a coboundary operator ($ \d^2 = 0 $)
A function $ \alpha_n $ which satisfies the equation $ \d \alpha_n = 0 $ is
called {\it n-cocycle}. A cocycle which can be presented inthe form $
\alpha_n = \d \beta $ is called {\it coboundary} or {\it trivial cocycle}.
We use the term trivial cocycle in a more narrow sense to denote a cocycle
which is a coboundary of a local functional. In our case the elements of $
M $ are the Yang-Mills fields and $ G $ is the group of gauge
transformations.  In this paper we presented a construction of 1-cocycle
on the gauge group (WZ action) and infinitesimal 2-cocycle (Schwinger term
in Gauss law commutator), now we shall show how these cocycles can be
constructed with the help of descent procedure.

We shall use the language of external forms. In this language the
Yang-Mills field is described by matrix valued 1-form $ A = \am dx^\mu $.
The Yang-Mills field strength is described by the 2-form $ F = dA + A^2 $,
the wedge product is assumed.

Led us consider the Minkovski space-time as embbedded into some higher
dimensional space. In this space we can write the closed gauge invariant
6-form
\be
\omega_{-1}=\frac {1}{48\pi^2} \tr F^3
\label{2.32}
\ee
This form is locally exact
\be
\omega_{-1}=d\omega_{0}
\ee
where
\be
\omega_{0}=\frac{1}{48\pi^2}\tr (F^2A-FA^3+\frac 25  A^5)+d\omega^\prime
\quad (mod\, Z)
\label{2.33}
\ee
where $ \omega^\prime $       is an arbitrary 4-form. The 5-form $
\omega_0 $ being integrated over a five dimensional manifold gives five
dimensional Chern-Simons action. Action of the coboundary operator $ \d $
which in this case coincides with gauge variation on $ \omega_0 $ gives a
closed 5-form
\be
d\d \omega_0=\d  d\omega_0=\d \omega_{-1}=0
\label{2.34}
\ee
the explicit form of $ \d \omega_0 $ is given by the equation
\bea
\d  \omega_0&=&\frac{1}{48\pi^2}\tr (dgg^{-1})^5+\frac{-i}{48\pi^2} d(\tr
[(AdA+dAA+A^3)dgg^{-1}-\nonumber \\
& &-\frac 12 Adgg^{-1}Adgg^{-1}-A(dgg^{-1})^3]+\d \omega^\prime (A))
\label{2.35}
\eea
{}From this explicit expression follows that the $ \d \omega_0 $ is locally
exact. Let us introduce the notation $ d^{-1} \kappa $
\be
\int_{M_4} d^{-1}\kappa=\frac{i}{240\pi^2} \int_{D_5} \kappa
\label{2.36}
\ee
where $ D_5 $ is a five dimensional disc with boundary $ M_4 $, the usual
Minkovski space. Then locally
\be
\d \omega_0=d\omega_1
\label{2.37}
\ee
where $ \omega_1 $ is defined up to a global exact form and a coboundary.
Integrating this form over the four dimensional space we get
\be
\frac{1}{2\pi}\alpha_1(A,g)=\int_{M_4} \omega_1(A,g)=\int_{D_5}d
\omega_1(A,g) =\int_{D_5}\d \omega_0(A,g)
\label{2.38}
\ee
The functional $ \alpha_1 $ is defined up to the coboundary $ \d \alpha_0
$
\be
\alpha_0(A)=\int_{M_4} \omega^\prime (A)
\ee
{}From eq.(\ref{2.38}) it follows that
\be
\d \alpha_1=0 \quad (mod\, 2\pi)
\label{2.39}
\ee
i.e. $ \alpha_1 $ is 1-cocycle.

If to choose $ \omega^\prime $ to be zero \cite{4} one gets the standard WZ
action which breaks the $ SO(N) $ invariance. The corresponding  $
\omega_0 $ is not $ SO(N) $ invariant for $ d \omega^\prime =0 $. Choosing
$ \omega^\prime $ in such a way to restore the $ SO(N) $ gauge invariance
of $ \omega_0 $ we shall get by the descent procedure the $ SO(N) $
invariant WZ action presented above. Using $ \d $ and $ d^{-1} $ to
continue the descent procedure we can calculate 2-cocycle. It is also
defined up to a coboundary and using this freedom one can get the
Schwinger term vanishing on $ SO(N) $ subgroup.

\section {APPENDIX}
\setcounter{equation}{0}

In this appendix we prove the gauge invariance of the integration measure
\be
DA_i DE_i D\phi (det\, M)^{1/2} \delta(C)
\label{b.1}
\ee
where $C$ is the secondary constraint
\be
C({\bf x})= \{C_p({\bf x}) e^p({\bf x}), H \} \sim \e^{pqrst}
R_p \Omega_{qr}({\bf x}) \Omega_{st} ({\bf y})
\label{b.2}
\ee
Where
\bea
R_p&=&\tr s_p f; \nonumber \\
\Omega_{pq}&=& \frac{i}{96 \pi^2} \e^{ijk} \tr ([s_p , s_q]f_1+
+s_p f_i s_q \widetilde{A}_k
-s_qf_is_p \widetilde{A}_k)
\label{b.3}
\eea
Explicite forms of $f$, $f_1$ and $f_i$ are given in the section 5.

$M$ is the matrix of Pisson brackets of the constraints
\be
M({\bf x},{\bf y})=\left(
\begin{array}{cc} \Omega_{pq} ({\bf x},{\bf y}) & v_{p}({\bf x},{\bf y})
\\ -v_{q}({\bf y},{\bf x}) & v({\bf x},{\bf y}) \end{array} \right)
\label{b.4}
\ee
where
\be
v_{p}({\bf x},{\bf y})=\{
C_{p}({\bf x}),C({\bf y})\} ,\qquad v({\bf x},{\bf y})=\{ C({\bf
x}),C({\bf y})\}
\label{b.5}
\ee
Firstly we prove that the gauge transformation of $C({\bf x})$ has a form
\be
C({\bf x}) \rightarrow  \det \left( \frac{ \p
\phi^p}{\p \widetilde{\phi^q}} \right) C({\bf x})
\label{b.6}
\ee
It follows from the explicit expressions for $f$, $f_1$ and $f_i$ and
$\widetilde{A_i}$ that under the gauge transformation they change as follows
\be
f_* \rightarrow g^{-1} f_* g
\label{b.7}
\ee
The law for $s_p(\phi )$ follows from the definition of $s_p $
\bea
\widetilde{s}_p(\widetilde{\phi })& = &\frac{\p s(\widetilde{\phi })}
{\p \widetilde{\phi }^p} s^{-1}(\widetilde{\phi })
=g^{-1}\frac{\p s(\phi )}
{\p \phi^p} s^{-1}(\phi )g=
\nonumber \\
&=&g^{-1}s_q(\phi )g \frac{\p \phi^q}{\p \widetilde{\phi}^p}
\label{b.8}
\eea
{}From eqs. (\ref{b.2}, \ref{b.3}, \ref{b.7}) we get the transformation law
for $C({\bf x}) $ \bea C({\bf x})& &\rightarrow \e^{p^\prime q^\prime
r^\prime s^\prime t^\prime} \frac{\p \phi^p}{\p
\widetilde{\phi}^{p^\prime}} \dots \frac{\p \phi^t}{\p
\widetilde{\phi}^{t^\prime}}R_p \Omega_{qr} \Omega_{st}= \nonumber \\ &=&
 det\, \left( \frac{\p \phi}{\p \widetilde{\phi}} \right) \e^{pqrst} R_p
\Omega_{qr} \Omega_{st}= det\, \left( \frac{\p \phi}{\p \widetilde{\phi}}
\right) C({\bf x}) \label{b.9} \eea To get the transformation law for
$det\, M$ we use the following equations \be v_p({\bf x,y}) \rightarrow
\frac{\p \phi^q}{\p \widetilde{\phi}^p}({\bf x}) v_q({\bf x,y}) \,det\,
\left( \frac{\p \phi}{\p \widetilde{\phi}} \right) ({\bf y}) \label{b.10}
\ee
\be
v({\bf
x,y}) \rightarrow \,det\, \left( \frac{\p \phi}{\p \widetilde{\phi}}
\right)({\bf x}) v({\bf x,y}) \,det\, \left( \frac{\p \phi}{\p
\widetilde{\phi}} \right)({\bf y})
\label{b.11}
\ee
which are valid on
the constraint surface $C=0$.  These equations lead to the following
transformation of the matrix $M$
\bea
M({\bf x},{\bf y}) &\rightarrow &
\left( \begin{array}{cc} \frac{\p \phi^p}{\p \widetilde{\phi}^r}({\bf x})
& 0 \\ 0 & det \left( \frac{\p \phi}{\p \widetilde{\phi}}({\bf x})\right)
\end{array} \right) \left( \begin{array}{cc} \Omega_{pq} ({\bf x},{\bf y})
& v_{p}({\bf x},{\bf y}) \\ -v_{q}({\bf y},{\bf x}) & v({\bf x},{\bf y})
\end{array} \right) \times
\nonumber \\
&& \quad
\left( \begin{array}{cc} \frac{\p \phi^q}{\p
\widetilde{\phi}^s}({\bf y}) & 0 \\
0 & det \left( \frac{\p \phi}{\p \widetilde{\phi}}({\bf y}) \right)
\end{array} \right)
\label{b.12}
\eea
and the $det\, M$ transforms as follows
\be
(detM)^{\frac{1}{2}} \rightarrow  \left(det \frac{\p \phi}{\p
\widetilde{\phi}}\right)^{2} (detM)^{\frac{1}{2}}
\label{b.13}
\ee
Let us firstly prove eq.(\ref{b.10}). The constraint $C$ has the following
structure
\be
C({\bf y}) = \tr E({\bf y})B({\bf y})
\label{b.14}
\ee
where $B({\bf y})$ depends on $s({\bf y})$, $\widetilde{A}_i ({\bf y})$ and
$s_p ({\bf y})$. Using the Leibnitz rule one can write the Poisson bracket
of $C_p({\bf x})$ and $ C({\bf y})$ in the form
\bea
v_p({\bf x, y})&=& \tr \{C_p({\bf x}), E_i({\bf y}) \} B_i({\bf y}) +
\nonumber \\
&+& \tr E_i({\bf y}) \{ C_p ({\bf x}), B({\bf y}) \}
\label{b.15}
\eea
The first term can be calculated using eq. (\ref{comm}). We get in this way
\bea
\{C_p({\bf x}), E_i^a({\bf y}) \}B_i & \sim & \epsilon^{ijk} \tr s_p(\{ \la
, F_{jk}-\frac12 s F_{jk}^T s^{-1} - \widetilde{A_j}  \widetilde{A_k}
\}- \nonumber \\
& - &  \widetilde{A_j} \la \widetilde{A_k} ) B_i \delta ({\bf x}-{\bf y})
\label{b.15a}
\eea
Using eqs. ({\ref{b.8}) and ({\ref{b.15a}) we see that this term is in
agreement with eq.(\ref{b.10}). To find the transformation law of the
second term we need the relations
\be
\{C_p({\bf x}), s({\bf y}) \}  \rightarrow \frac{\p \phi^q}{\p
\widetilde{\phi}^p }({\bf x}) g^{-1}({\bf y})\{C_q({\bf x}), s({\bf y})
\} g^{-1,T}({\bf y})
\label{b.16}
\ee
\be
\{C_p({\bf x}), \widetilde{A}_i \} \rightarrow  \frac{\p \phi^q}{\p
\widetilde{\phi}^p }({\bf x}) g^{-1}({\bf y})\{C_p({\bf x}),
\widetilde{A}_i  \}  g({\bf y})
\label{b.17}
\ee
\bea
\{C_p({\bf x}), s_q({\bf y}) \} & \rightarrow &
\frac{\p \phi^{p_1}}{\p \widetilde{\phi}^p }({\bf x})
\frac{\p \phi^{q_1}}{\p \widetilde{\phi}^q }({\bf y}) g^{-1}({\bf
y}) \times \nonumber \\
& \times & \{C_{p_1}({\bf x}), s_{q_1}({\bf y}) \} g^{-1,T}({\bf y})-
g^{-1}s_{q_1}g \frac{\p \phi^{p_1}}{\p \widetilde{\phi}^p }
\frac{\p^2 \phi^{q_1}}{\p \widetilde{\phi}^{p_1} \p \widetilde{\phi}^q }
\delta ({\bf x-y})
\label{b.18}
\eea
To prove eqs.(\ref{b.16}--\ref{b.18}) we note that $C_p$ can be presented
in the form
\be
C_p ({\bf x})=\pi_p+\varphi_p(s,A)
\label{b.19}
\ee
where $\pi_p$ is the canonical momentum for $\phi^p$. The second term in
eq.(\ref{b.19}) depends only on $s$ and $A_i$ and therefore commutes with
$s$ and $A_i$. The Poisson bracket  of $\pi_p$ with any functional $W$ of
$\phi^p$ is equal to
\be
\{\pi_p ({\bf x}), W(\phi ) \} = -\frac{\d }{\d \phi^p ({\bf x})}W(\phi )
\label{b.20}
\ee

If the last term in eq. (\ref{b.18}) were absent one would get
for $v_p$ the transformation law ({\ref{b.10}). This term leads to an
additional contribution in the transformation law for $v_p({\bf x})$.
However one can show that this contribution vanishes on the constraint
surface $C=0$. To prove it let us note that the constraint $C$ can be
represented as follows
\be C({\bf x})=\tr \left( \e^{pqrst} (s_p \otimes
s_q \otimes s_r \otimes s_s \otimes s_t ) V({\bf x}) \right)
\label{b.21}
\ee
where $V({\bf x})$ does not depend on $s_p$. Thus the additional term
is equal to
\bea
\Delta &=& \tr [\e^{uqrst}(s_{p_1qrst}- s_{qp_1rst}+
s_{qrp_1st} -s_{qrsp_1t}+ \nonumber \\
&+& s_{p_1qrst}) \frac{\p^2
\phi^{p_1}}{\p \widetilde{\phi}^{q_1} \p \widetilde{\phi}^u }\frac{\p
\phi^{q_1}}{\p \widetilde{\phi}^{p}} V({\bf x}) ] \delta ({\bf x-y})
\label{b.22}
\eea
where $s_{pqrst} = s_p \otimes s_q \otimes s_r \otimes s_s \otimes s_t $.
The combination in the curly brackets is proportional to
\be
\e^{uqrst}(s_{p_1qrst}- \ldots ) \sim {\delta_{p_1}}^u \e^{wqrst} s_{wqrst}
\label{b.23}
\ee
It is obvios from eq.(\ref{b.23}) that $\Delta $ is proportional to $C$.
Therefore we proved the validity of eq.({\ref{b.10}). Eq. (\ref{b.11}) can
be derived in a similar way. Taking into account that $DA$ and $DE$ are
invariant and $D\phi $ transforms as follows
\be
D\phi \rightarrow det \, \left( \frac{\p \phi}{\p \widetilde{\phi }}
\right) D \widetilde{\phi }
\label{thelast}
\ee
we see that the
integration measure (\ref{b.1}) is gauge invariant.

\end{document}